\documentclass[cameraready]{Interspeech}

\title{Beyond Acoustic Sparsity and Linguistic Bias: A Prompt-Free Paradigm for Mispronunciation Detection and Diagnosis}

\author[affiliation={1}, orcid=0009-0006-1975-8956]{Haopeng}{Geng}
\author[affiliation={1}, orcid=0000-0002-5079-3091]{Longfei}{Yang}
\author[affiliation={1}, orcid=0009-0004-8223-9716]{Xi}{Chen}
\author[affiliation={1}, orcid=0009-0008-5134-1936]{Haitong}{Sun}
\author[affiliation={1}, orcid=0000-0002-6265-9674]{Daisuke}{Saito}
\author[affiliation={1}, orcid=0000-0002-8778-9555]{Nobuaki}{Minematsu}

\address{
    $^1$ Graduate School of Engineering, The University of Tokyo, Japan 
}

\email{kevingenghaopeng, chenxi, longfei, sunhaitong, dsk\_saito, mine@gavo.u-tokyo.ac.jp}

\usepackage{seqsplit} 
\usepackage[table]{xcolor} 
\usepackage{comment}
\usepackage{amsmath}
\usepackage{hyperref}
\usepackage{booktabs}
\usepackage{multirow}
\usepackage{algorithmic}
\usepackage{graphicx}
\usepackage{textcomp}
\usepackage{xcolor}
\usepackage{siunitx}
\usepackage{makecell}
\usepackage{caption}
\usepackage{mathrsfs}
\usepackage{bm}      
\usepackage{subcaption}  
\usepackage{bbding}
\usepackage{cite}
\usepackage{tipa}
\begin{document}

\maketitle 



\begin{abstract}

Mispronunciation Detection and Diagnosis (MDD) requires modeling fine-grained acoustic deviations. However, current ASR-derived MDD systems often face inherent limitations. In particular, CTC-based models favor sequence-level alignments that neglect transient mispronunciation cues, while explicit canonical priors bias predictions toward intended targets. To address these bottlenecks, we propose a prompt-free framework decoupling acoustic fidelity from canonical guidance. First, we introduce CROTTC, an acoustic model enforcing monotonic, frame-level alignment to accurately capture pronunciation deviations. Second, we implicitly inject mispronunciation information via the IF strategy under the knowledge transfer principle. Experiments show CROTTC-IF achieves a 71.77\% F1-score on L2-ARCTIC and 71.70\% F1-score on the Iqra'Eval2 leaderboard. With empirical analysis, we demonstrate that decoupling acoustics from explicit priors provides highly robust MDD\footnote{Source code is available at: \url{https://github.com/Secondtonumb/IF-MDD}}.
\end{abstract}

\section{Introduction}
\label{sec:intro}
Mispronunciation Detection and Diagnosis (MDD) plays an indispensable role across a spectrum of applications, ranging from general Computer-Aided Pronunciation Training (CAPT) for L2 learners to religious domains such as Qur'anic recitation, as highlighted by the Iqra'Eval Challenges\cite{el2025iqra, kheir2025unifiedbenchmarkarabicpronunciation}. 
The core challenge of MDD lies in acoustic fidelity: transcribing speech exactly as realized to pinpoint deviations from canonical norms. Historically, MDD was formulated as a downstream application of Automatic Speech Recognition (ASR). Conventional systems typically used ASR features to generate Goodness-of-Pronunciation (GOP) scores \cite{witt2000phone} or relied on Extended Recognition Networks (ERNs) to catch predefined error patterns \cite{harrison2009implementation, li2016mispronunciation}. In this classical paradigm, MDD was essentially an ASR by-product.

With the rapid advancement of end-to-end (E2E) architectures and self-supervised learning (SSL) speech representations, the paradigm has shifted. MDD is now a standalone task, directly optimized for fine-grained phoneme diagnosis. However, despite this structural independence, MDD research inevitably remains under the heavy methodological shadow of ASR. While general ASR focuses on deducing semantic intent of a speaker despite acoustic imperfections, MDD demands objective, fine-grained acoustic fidelity. By uncritically adopting ASR's acoustic and linguistic optimization strategies, we believe that current MDD studies frequently fall into two fatal traps.

The first is the \textit{Acoustic Trap}, driven by the direct adoption of Connectionist Temporal Classification (CTC) \cite{graves2006connectionist}. As a de facto standard for E2E ASR, CTC usually serves as an acoustic model (AM) by maximizing the marginal probability of all valid alignment paths, forcing the model to prioritize global sequence correctness. However, when ported to the MDD task, this global optimization actively smooths over the subtle, transient acoustic variations (e.g., co-articulation, brief onset substitutions).
In doing so, CTC might erase the exact phonetic evidence that is critical for mispronunciation diagnosis. 

The second is the \textit{Linguistic Trap}, driven by the reliance on explicit canonical prompts or strong language model (LM) post-processing. In typical ASR systems such as CTC/AED\cite{watanabe2017hybrid} or RNN-T\cite{graves2012sequence}, LMs are powerful components to correct acoustic errors. By adjusting the AM/LM weight, ASR models can shift their preference to favor either acoustic or linguistic features.
In MDD, however, when models are guided by canonical priors, they naturally exhibit an over-correction tendency. Instead of objectively diagnosing the actual phonetic deviations from the acoustic signal, the system defaults to the linguistically most likely sequence, which is almost always the canonical text. 

To break free from ASR's legacy and overcome these bottlenecks, we propose \textbf{CROTTC-IF}, a unified, canonical prompt-free paradigm, designed exclusively for the standalone MDD task. Our main contributions are summarized as follows:
\begin{itemize}
\item \textbf{Frame-wise Acoustic Modeling (CROTTC):} We introduce \textbf{C}onsistency-\textbf{R}egularized \textbf{O}ptimal \textbf{T}emporal \textbf{T}ransport \textbf{C}lassification. Unlike standard CTC, this acoustic model enforces strict monotonic, frame-wise alignments, effectively mitigating inherent sparsity and delayed emission issues to preserve fine-grained, transient mispronunciation cues.

\item \textbf{Knowledge Transfer via Indirect Fusion (IF):} Grounded in the Learning Using Privileged Information (LUPI) paradigm, we introduce \textbf{I}ndirect \textbf{F}usion stragety to implicitly integrate linguistic priors. By treating canonical texts and mispronunciation patterns as privileged data during training, IF provides soft linguistic guidance to the language model, preventing the override of faithful acoustic evidence during inference.

\item \textbf{Comprehensive Analysis of Canonical Bias:} With delicately designed Multi-modality Large Language Models (LLMs) and various prompt templates, we quantitatively investigate the impact of explicit canonical prior in MDD. Our findings demonstrated that over-reliance on canonical priors can degrade detection sensitivity, highlighting the necessity of balancing linguistic context with acoustic fidelity.
\item \textbf{State-of-the-Art Performance:} Operating entirely without auxiliary data or explicit canonical prompts, the CROTTC-IF framework achieves highly competitive results. It demonstrates strong generalization across diverse benchmarks, ranging from general L2 English corpora (L2-ARCTIC, ERJ, speechocean762) to the specialized Arabic Qur'anic recitation task (Iqra'Eval2).
\end{itemize}

\section{Related Works}

\begin{figure*}[t]
	\centering
	\includegraphics[width=0.9\textwidth]{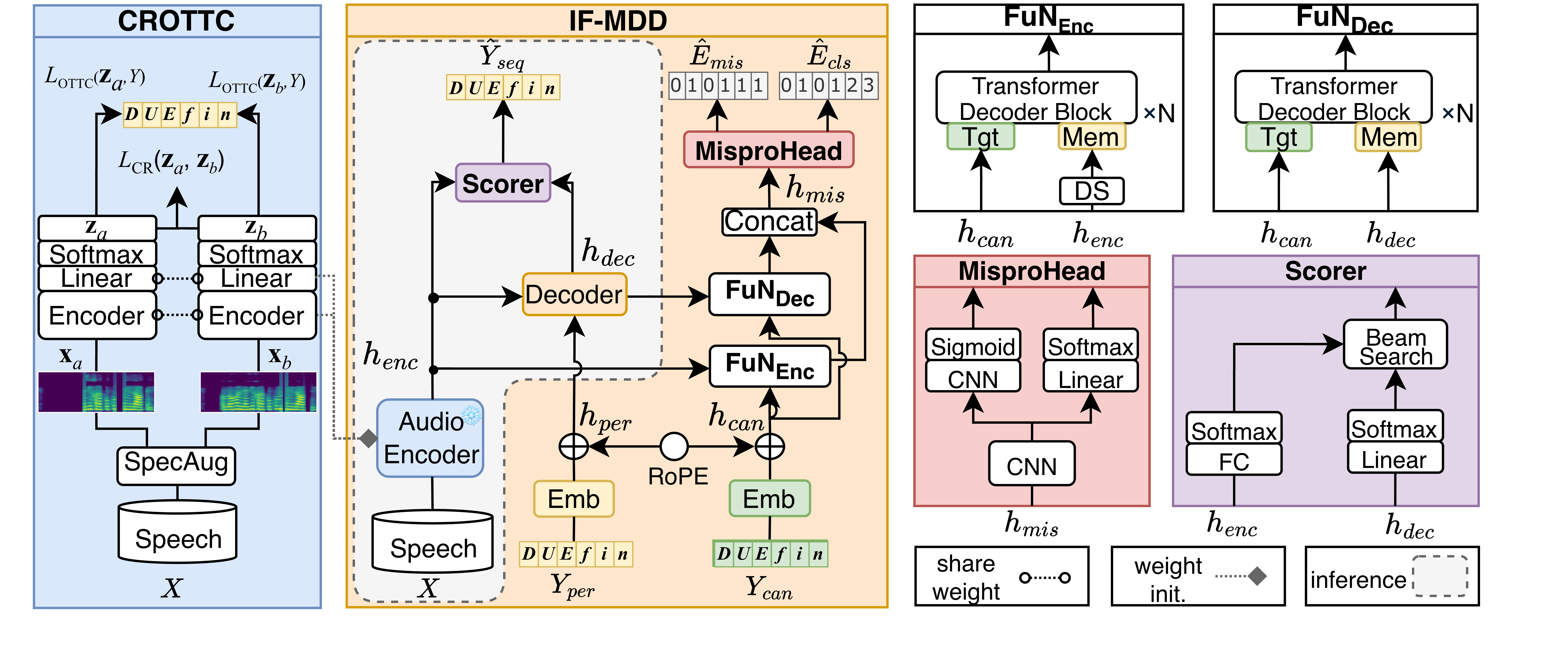}
\vspace{-7mm}
\caption{Overview of the CROTTC-IF architecture. From left to right, it comprises the CROTTC AM (blue), the lightweight IF-MDD LM (orange), and the detailed IF components.}
	\label{fig:iqra_system}
\vspace{-6mm}
\end{figure*}

Currently, modern MDD approaches broadly fall into two categories: dictation-style and text-prompting style.

\subsection{Dictation-Style MDD and the Acoustic Trap}

Dictation-style methods aim to recognize the uttered phoneme sequence exclusively from acoustic-related features. For example, Leung et al. introduced a CTC framework and demonstrated the feasibility of capturing L2 phoneme information from acoustic features alone \cite{cnn-rnn-ctc}. More recent studies have leveraged self-supervised learning (SSL) speech representations, such as wav2vec 2.0~\cite{wav2vec2}, HuBERT~\cite{hsu2021hubert, boito2024mhubert} and WavLM~\cite{chen2022wavlm}, which provide robust phonemic context features. For instance, \cite{peng2021study} employed a fine-tuned wav2vec 2.0 model with a CTC decoder, achieving promising diagnostic accuracy. To strengthen free phoneme recognition, researchers have also incorporated semi-supervised pseudo-labeling~\cite{yang2022improving}, dual-path SSL for predicting manner-of-articulation~\cite{el2023multi}, and acoustic-to-articulatory inversion leveraging electromagnetic articulography (EMA) data~\cite{lin2024pg}. 

This dictation-style MDD also dominated the recent Iqra'Eval 2025 challenge \cite{el2025iqra}. For instance, \textit{Hafs2Vec}~\cite{ibrahim-2025-hafs2vec} utilized extensive professional reciter data, while \textit{Metapseud}~\cite{mansour-2025-metapseud} applied curriculum learning and beam-search decoding. Despite these feature and data enhancements, the fundamental reliance on CTC decoders introduces a potential vulnerability: the \textit{Acoustic Trap}\footnote{Further detailed in Sec.~\ref{sec:ctc_limit}}.

Conversely, frame-wise alignment methods have demonstrated greater potential for MDD. For example, Feng et al. pioneered the use of attention-based sequence labeling to map individual acoustic frames directly to phone labels\cite{9052975}.
Similarly, Lin et al. modeled MDD as a sequence labeling task via a joint-alignment mechanism \cite{lin2022phoneme}. More recently, Tu et al. introduced a retrieval-based framework for frame-wise classification, achieving competitive performance without task-specific training \cite{tu2025mispronunciation}. 
These advancements indicate that leveraging \textbf{dense, frame-wise representations} provides superior performance, outperforming sparse features for robust MDD.

\subsection{Text-Prompting MDD and the Linguistic Trap}

In reading-aloud scenarios, canonical phoneme sequences are inherently available. Consequently, integrating this prior knowledge into MDD models has become another standard practice. For instance, Ye et al. proposed an approach to leverage acoustic, phonetic and linguistic embeddings for MDD\cite{ye2022approach}. Peng et al. employed a gating mechanism and contrastive loss to dynamically regulate canonical embeddings~\cite{peng2022text}. Zheng et al. utilized coupled-cross attention for explicit acoustic-canonical fusion~\cite{zheng2022coca}. Similarly, Yan et al. injected implicit linguistic priors using phoneme lookup tables and graph networks to model frequent mispronunciation patterns~\cite{yan2023effective, peppanet}.

Entering the era of Large Language Models (LLMs), recent studies have shifted toward leveraging massive linguistic priors for posterior correction in MDD.
Current methods range from adapting LLM-based ASR architectures~\cite{ma2026slam} to fine-tuning native SpeechLM~\cite{Qwen2-Audio}. For instance, Wu et al. adapted LLMs for MDD by coupling a pretrained acoustic front-end with a trainable projector and LLM backbone~\cite{wu2024prompting}. Other researchers adapted multi-modality LLM to embed canonical sequences and predefined error patterns directly into the latent space~\cite{wu2025integrating, Song2025PhonemeControlledLW}. Beyond simple transcription, LLM reasoning is also deployed to generate educational feedback, offering nuanced diagnostics alongside canonical norms~\cite{zhong2024leveraging, xie2026mispronunciation}.

Despite reported successes, the practical superiority of text-prompted and LLM-assisted methods remains debatable. Many systems are evaluated on private datasets, precluding reproducible comparisons. And on open benchmarks, they frequently underperform compared to conventional models that rely solely on acoustic modeling.

Beyond evaluation discrepancies, we believe that some of these canonical-dependent methods inherently fall into the \textit{Linguistic Trap} due to a shared flaw: \textbf{canonical information leakage}. Whether by explicitly including canonical information as part of the input, or by using a strong language model as the model's prior knowledge, these approaches can all be categorized as suffering from canonical information leakage. Although many models carefully design their canonical information usage, canonical information can easily override the subtle acoustic information. Consequently, the model becomes overly tolerant of errors, masking critical acoustic deviations and sacrificing diagnostic objectivity. Furthermore, requiring explicit canonical texts during inference renders these methods entirely impractical for spontaneous speech or shadowing scenarios~\cite{mine22_SP, geng2024APSIPA, geng2025perception}, restricting the real-world applicability of MDD.

\section{Consistency Regularization on Optimal Temporal Transport Classification}
\label{sec:CROTTC}
In this section, we analyze the inherent limitations of CTC and introduce the architecture of our proposed frame-wise dense acoustic model: \textbf{CROTTC}. Our approach, as illustrated in the left panel of Figure \ref{fig:iqra_system}, is built upon two core pillars: Consistency Regularization \textbf{(CR)} and Optimal Temporal Transport Classification\textbf{ (OTTC)}.

\subsection{Limitations of Connectionist Temporal Classification}
\label{sec:ctc_limit}
 
Connectionist Temporal Classification (CTC) \cite{graves2006connectionist} has established itself as a fundamental criterion for sequence-to-sequence modeling, particularly in the ASR domain. By introducing a blank token and optimizing the marginal probability over all valid alignments, CTC enables end-to-end training without requiring explicit frame-level annotations for target units (e.g., phonemes or subwords). Consequently, the model tends to produce a highly sparse posterior distribution, concentrating probability mass on a few highly discriminative frames \cite{zeyer2021does}. Moreover, these non-blank emission peaks are frequently delayed, shifting towards the tail end of their corresponding acoustic segments—a phenomenon widely recognized as CTC's delayed behavior \cite{tian2022bayes, yao2023delay}.

This sparsity and delayed behavior pose a significant challenge for precise segment detection tasks. In practice, a mispronounced phoneme in MDD does not necessarily manifest as a complete substitution starting from the very first frame; rather, it often involves co-articulation errors or a partial blend of canonical and abnormal acoustic traits. To illustrate, consider a speaker attempting to produce the diphthong \textipa{/aI/} (as in ``buy'') but erroneously substituting the initial vowel element to produce \textipa{/OI/} (as in ``boy''). Acoustically, both diphthongs share an identical trailing off-glide (\textipa{/I/}). Driven by CTC's inherent delayed emission tendency, the model will typically align the probability peak for the target label (\textipa{/aI/}) exclusively with this shared off-glide. Consequently, the preceding frames containing the crucial evidence of the mispronunciation (\textipa{/O/}) can be masked by blank tokens and entirely ignored. While this highly discriminative frame selection is beneficial for standard ASR, it discards the fine-grained acoustic cues that are essential for sensitive diagnostic tasks like MDD.

\subsection{Optimal Temporal Transport Classification}
\label{sec:ottc}
To address the inherent shortcomings of CTC, we leverage Optimal Temporal Transport Classification (OTTC)~\cite{kaloga2025ottc}, a differentiable framework rooted in one-dimensional optimal transport that enables dense, frame-level alignments.
 
\subsubsection{Sequence Alignment via Optimal Transport}
Let $\mathbf{X} = \{x_i\}_{i=1}^n$ denote the acoustic frame sequence and $\mathbf{Y} = \{y_j\}_{j=1}^m$ the perceived phoneme sequence. We define that the alignment between $\mathbf{X}$ and $\mathbf{Y}$ can be solved with an optimal transport plan $\gamma \in \mathbb{R}^{n \times m}_+$, which serves as a monotonic mapping governing the frame-to-label correspondence. In optimal transport theory,  $\gamma$ is parameterized by two discrete distributions: the frame weights $\alpha \in \Delta^n$ derived from $\mathbf{X}$, and the label weights $\beta \in \Delta^m$ associated with $\mathbf{Y}$, where $\Delta^k = \{v \in \mathbb{R}^k_+ \mid \sum_{i=1}^k v_i = 1\}$ is the probability simplex. Here, the frame weights $\alpha$ can be predicted from the input via a neural network $W$:
\vspace{-0.50\baselineskip} 
\begin{equation}
    \alpha[\mathbf{X}, W] = \mathrm{softmax}(W(x_1), \ldots, W(x_n))^\top,
    \label{eq:alpha}
\end{equation}
and $\beta$ is typically fixed as a uniform distribution to evenly assign each target token's contribution. Subsequently, the alignment plan $\gamma(\alpha, \beta)$ is computed as the solution to the 1D optimal transport problem:
\vspace{-0.50\baselineskip} 
\begin{equation}
    \gamma(\alpha, \beta) = \arg\min_{\gamma \in \Gamma_{\alpha,\beta}} \sum_{i=1}^n \sum_{j=1}^m \gamma_{i,j} \, \lVert i - j \rVert_2^2,
    \label{eq:ot_plan}
\end{equation}
where $\Gamma_{\alpha,\beta} = \left\{ \gamma \in \mathbb{R}_+^{n \times m} \mid \gamma \mathbf{1}_m = \alpha, \, \gamma^\top \mathbf{1}_n = \beta \right\}$ denotes the set of valid couplings preserving the marginal distributions. 
Based on the \textit{Sequence Optimal Transport Distance (SOTD)} framework~\cite{kaloga2025ottc}, which defines a pseudo-metric over sequence space, we derive the OTTC loss as the expected transport cost under the cross-entropy metric:
\vspace{-0.50\baselineskip} 
\begin{equation}
    \mathcal{L}_{\text{OTTC}} = -\sum_{i=1}^{n} \sum_{j=1}^{m} \gamma_{i,j}(\alpha, \beta) \cdot \log  p(y_j \mid x_i),
    \label{eq:ottc_loss}
\end{equation}
where $p(y_j \mid x_i)$ is the posterior probability. Since $\alpha$ is parameterized via a neural network $W$ (Eq.~\ref{eq:alpha}) and $\beta$ is fixed, the model parameters can be optimized in an end-to-end manner by minimizing $\mathcal{L}_{\text{OTTC}}$.

\noindent\textbf{Comparison to CTC.} 
\begin{figure}[t]
	\centering
	\includegraphics[width=0.45\textwidth]{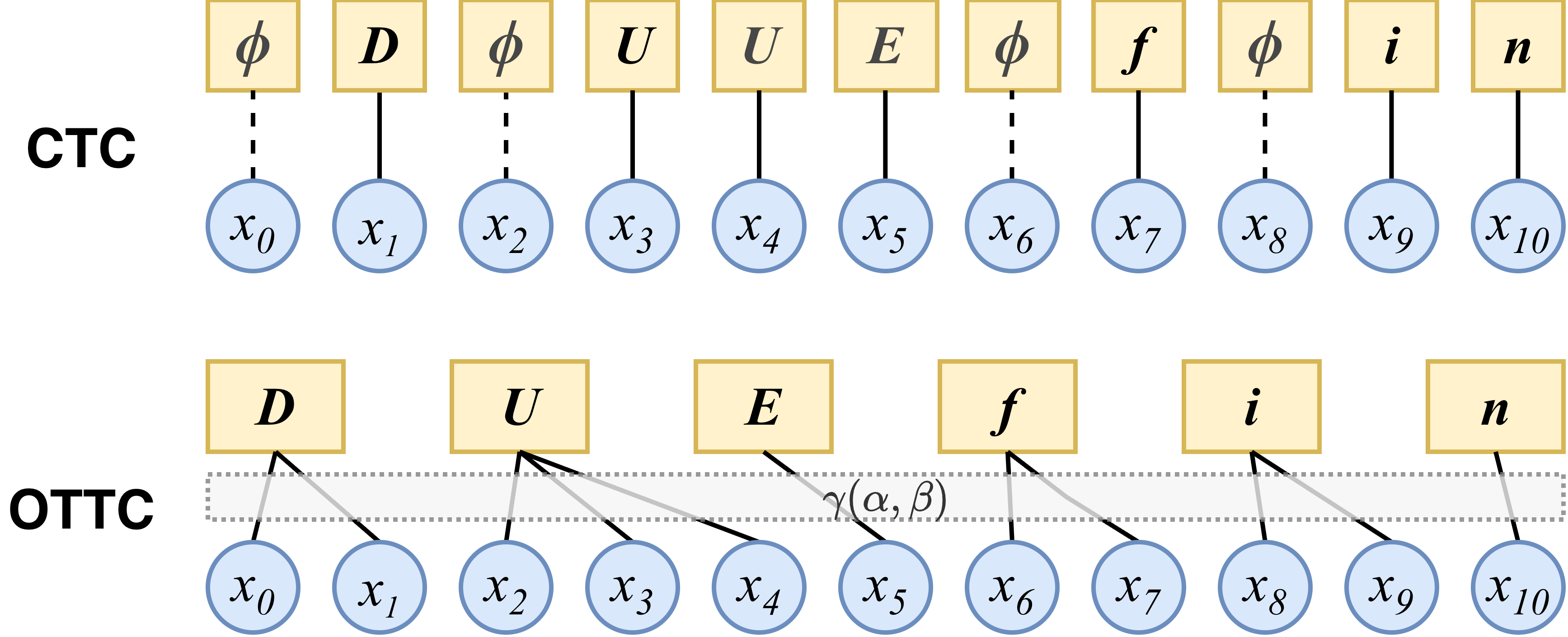}
    \vspace{-2mm}
        \caption{Comparison between CTC and OTTC, where $\phi$ is the blank token and $\gamma(\alpha,\beta)$ is the optimal transport plan.}
	\label{fig:CTC_OTTC_compare}
\vspace{-3mm}
\end{figure}
Fig.~\ref{fig:CTC_OTTC_compare} illustrates the fundamental distinction in alignment strategies. 
Specifically, CTC maximizes the marginal probability over \textit{all} valid monotonic paths:
\vspace{-0.50\baselineskip} 
\begin{equation}
    \mathcal{L}_{\text{CTC}} = -\log \sum_{\pi \in \mathcal{B}^{-1}(\mathbf{Y})} \prod_{i=1}^n p(\pi_i \mid x_i),
    \label{eq:ctc_loss}
\end{equation}
where $\mathcal{B}^{-1}(\mathbf{Y})$ denotes the set of all paths collapsed to $\mathbf{Y}$. 
In contrast, OTTC maximizes the probability of a \textit{single} optimal monotonic path $\gamma(\alpha, \beta)$ via Eq.~\eqref{eq:ot_plan} and~\eqref{eq:ottc_loss}. 
This formulation yields dense, frame-level alignments without blank-token dominance, as $\gamma$ explicitly models the frame-to-label correspondence rather than marginalizing over sparse paths.
 
\subsection{Consistency Regularization} 
\label{sec:cr}
While OTTC captures fine-grained mispronunciation cues, acoustic models can be over-sensitive to local variations, leading to spurious insertions. To mitigate this, we also introduce \textit{Consistency Regularization (CR)} into our training pipeline\cite{yaocr}. Specifically, as shown in Fig~\ref{fig:iqra_system}, for each input utterance $\mathbf{X}$, we generate two augmented views $\mathbf{X}_a$ and $\mathbf{X}_b$ using stochastic perturbations\footnote{See Sec.~\ref{sec:training_config} for spectrogram augmentation details.}. Let $\mathbf{Z}_a$ and $\mathbf{Z}_b$ denote the posterior probability distributions output by the model for each view, where $\mathbf{Z}^{(i)} \in \Delta^K$ represents the probability vector at frame $i$ (equivalent to $p(y \mid x_i)$ in Eq.~\ref{eq:ottc_loss}). 
The consistency regularization loss is defined as the symmetric Kullback-Leibler (KL) divergence between these distributions:
\vspace{-0.55\baselineskip} 
\begin{equation}
    \mathcal{L}_{\text{CR}} = \frac{1}{2n} \sum_{i=1}^{n} \left( \mathrm{KL}\left(\mathbf{Z}_a^{(i)} || \mathbf{Z}_b^{(i)}\right) + \mathrm{KL}\left(\mathbf{Z}_b^{(i)} || \mathbf{Z}_a^{(i)}\right) \right),
    \label{eq:cr_loss}
\end{equation}
where $n$ is the number of frames. By minimizing the distribution distance and maximizing the mutual information between the two branches, CR enables each branch to receive supervision from the other, which encourages the shared encoder to perform self-distillation on frame-level posterior distributions. 
Additionally, relatively long-term time masking implicitly encourages \textit{masked language modeling}-like behavior: the model learns to reconstruct occluded acoustic frames by leveraging contextual information from unmasked regions. Together, these mechanisms stabilize frame-level predictions and reduce sensitivity to local acoustic noise. Consequently, the training criteria for CROTTC is defined as:
\vspace{-0.55\baselineskip} 
\begin{equation}
    \mathcal{L}_{\text{AM}} = \mathcal{L}_{\text{CR}} + \eta \left( \mathcal{L}_{\text{OTTC}}(\mathbf{Z}_a, \mathbf{Y}) + \mathcal{L}_{\text{OTTC}}(\mathbf{Z}_b, \mathbf{Y}) \right),
    \label{eq:total_loss}
\end{equation}
where hyperparameter $\eta$ is set to 1.0.

\section{Indirect Fusion of Mispronunciation Information via Knowledge Transfer}
\label{sec:IFMDD}
In this section, we detail our Indirect Fusion (IF) strategy for incorporating mispronunciation information into LM, named\textbf{ IF-MDD}. As illustrated in the middle and right panels of Fig. \ref{fig:iqra_system}, IF-MDD treats canonical phonemes and mispronunciation cues as privileged information available exclusively during the training phase. These cues are leveraged to guide the model’s latent representations via backpropagation. 

\subsection{Learning using Privilege Information}
\label{sec:LUPI}
Learning using Privileged Information (LUPI) is a training paradigm that fundamentally operates as a knowledge transfer mechanism, leveraging additional information available exclusively during the training phase to implicitly guide the model \cite{vapnik2009new, vapnik2015learning}. While this paradigm has already been successfully applied to various speech-related tasks \cite{fukuda2020implicit, fukuda2017efficient, 10832134}, its most compelling advantage lies in its effectiveness when training data is limited or labels are scarce \cite{lopez2015unifying, 10.5555/3692070.3694170, ortiz2023does}. Given the inherent data scarcity of expert-annotated MDD corpora, adopting the LUPI framework is a natural and highly principled choice.

\subsection{Indirect Fusion of Mispronunciation Information via Knowledge Transfer}
As illustrated in Fig.~\ref{fig:iqra_system}, IF-MDD consists of two main components: 1) an encoder--decoder backbone; 2) an auxiliary mispronunciation-aware teacher network. 


\begin{figure}[t]
	\centering
	\includegraphics[width=0.48\textwidth]{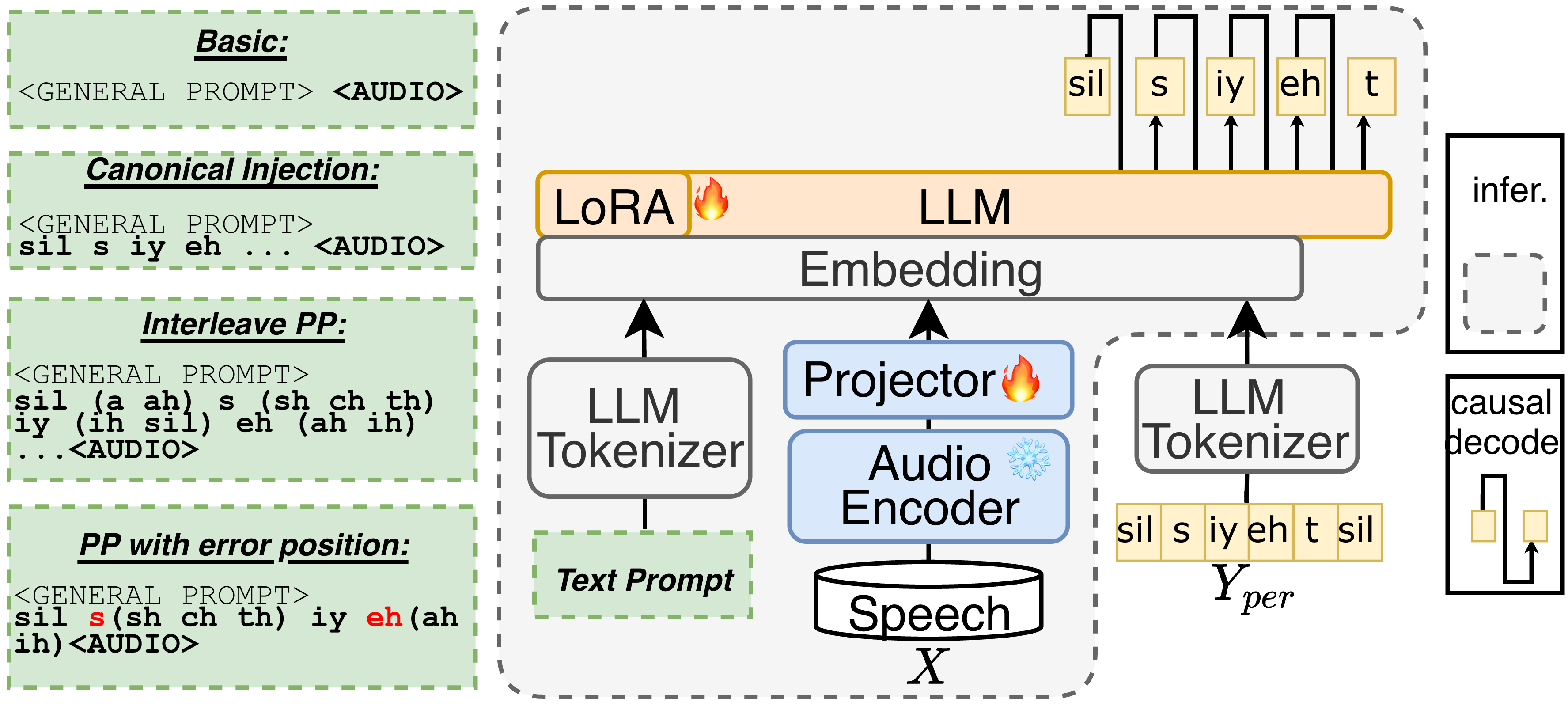}
    \vspace{-5mm}
        \caption{Illustration of LLM-based MDD, with alternative prompts.}
	\label{fig:LM2LLM}
\vspace{-5mm}
\end{figure}

\subsubsection{Encoder-decoder backbone} 
We employ an encoder--decoder architecture serving as the primary backbone of IF-MDD. Let the input waveform be $\mathbf{X} = \{x_i\}_{i=1}^n$ and the perceived phoneme sequence be $\mathbf{Y}_{\text{per}} = \{y^{\text{per}}_j\}_{j=1}^m$, where $n$ and $m$ denote the lengths of the speech signal and phoneme sequence, respectively. The backbone is formulated as:
\vspace{-0.5\baselineskip} 
\begin{align}
\mathbf{h}_{\text{enc}} &= \text{Enc}(\text{RoPE}(\mathbf{X})),\\
\mathbf{h}_{\text{per}} &= \text{RoPE}(\mathbf{Y}_{\text{per}}),\\
\mathbf{h}_{\text{dec}} &= \text{Dec}(\mathbf{h}_{\text{per}},\, \mathbf{h}_{\text{enc}}),
\end{align}
where $\text{Enc}(\cdot)$ and $\text{Dec}(\cdot)$ represent the encoder (AM) and decoder (LM), respectively. Rotary position embeddings $\text{RoPE}(\cdot)$ are applied to both $\mathbf{h}_{\text{enc}} \in \mathbb{R}^{n \times D}$ and $\mathbf{h}_{\text{dec}} \in \mathbb{R}^{m \times D}$ to enhance positional modeling \cite{su2024roformer}.

\subsubsection{Auxiliary mispronunciation-detection teacher network} 
\label{sec:teacher}
To explicitly model the mispronunciation information indicated by the misalignment between acoustic and canonical features, we introduce an auxiliary teacher network. This module leverages the canonical phoneme sequence $\mathbf{Y}_{\text{can}} = \{y^{\text{can}}_j\}_{j=1}^m$ and the corresponding phoneme-level ground-truth error sequence $\mathbf{E} = \{e_j\}_{j=1}^m$ as privileged information. It consists of two sequential stages: feature fusion and comprehensive error detection.

First, two parallel fusion networks ($\text{FuN}$) are employed to model multi-aspect feature interactions:
\vspace{-0.50\baselineskip} 
\begin{align}
  \mathbf{h}_{\text{can}} &= \text{RoPE}(\mathbf{Y}_{\text{can}}), \\
  \mathbf{h}_{\text{mis}}^{\text{enc}} &= \text{FuN}_{\text{enc}}(\mathbf{h}_{\text{can}},\, \text{DS}(\mathbf{h}_{\text{enc}})), \\
  \mathbf{h}_{\text{mis}}^{\text{dec}} &= \text{FuN}_{\text{dec}}(\mathbf{h}_{\text{can}},\, \mathbf{h}_{\text{dec}}), \\
  \mathbf{h}_{\text{mis}} &= \text{Concat}(\mathbf{h}_{\text{mis}}^{\text{enc}},\, \mathbf{h}_{\text{mis}}^{\text{dec}}),
\end{align}
where $\text{FuN}(\cdot)$ is composed of a transformer decoder structure and $\mathbf{h}_{\text{mis}} \in \mathbb{R}^{m \times D}$ denotes the fused representation. In both branches, $\text{FuN}(\cdot)$ uses $\mathbf{h}_{\text{can}}$ as the query and $\mathbf{h}_{\text{enc}}$ or $\mathbf{h}_{\text{dec}}$ as the memory. Notably, a downsampler $\text{DS}(\cdot)$ is applied within $\text{FuN}_{\text{enc}}(\cdot)$ to improve alignment.

Subsequently, our teacher network employs a dual-path detection sub-network to provide a comprehensive diagnosis. Specifically, we apply a shared CNN trunk on the fused representation $\mathbf{h}_{\text{mis}}$, followed by two task-specific projection heads:
\begin{align}
\mathbf{u} &= \text{CNN}(\mathbf{h}_{\text{mis}}),\\
\hat{E}_{\text{mis}} &= \sigma(\text{CNN}_{\text{bin}}(\mathbf{u})),\\
\hat{E}_{\text{cls}} &= \text{Softmax}(\text{Linear}_{\text{cls}}(\mathbf{u})),
\end{align}
where $\hat{E}_{\text{mis}}$ and $\hat{E}_{\text{cls}}$ are the predicted counterparts for the position and classification facets of $\mathbf{E}$, respectively. Here, $\hat{E}_{\text{mis}} \in [0, 1]^m$ indicates the probability of mispronunciation, while $\hat{E}_{\text{cls}}$ denotes the distribution over \{\textit{correct}, \textit{substitution}, \textit{deletion}, \textit{insertion\}}.

Crucially, this architectural design naturally facilitates knowledge transfer. Since the auxiliary mispronunciation heads are relatively lightweight and receive direct supervision from $\mathbf{E}$, they converge significantly faster than the primary sequence-to-sequence backbone. During training, this fast-converging teacher provides a strong, discriminative gradient signal that backpropagates through the fusion networks into both $\mathbf{h}_{\text{enc}}$ and $\mathbf{h}_{\text{dec}}$. Consequently, the diagnostic features from the teacher network act as a catalyst, helping the deeper backbone to converge effectively and internalize the mispronunciation-aware representations required for prompt-free inference.

\subsection{Inference and Training Criteria}
\label{sec:loss}
During inference, the auxiliary teacher is discarded. A scorer searches for the optimal hypothesis $\hat{\mathbf{Y}}$ by maximizing the interpolated log-probabilities between the AM and LM:
\begin{equation}
\begin{aligned}
    \hat{\mathbf{Y}} = \arg\max_{\mathbf{Y}} \big\{ & \lambda \log P_{\text{AM}}(\mathbf{Y} | \mathbf{h}_{\text{enc}}) \\
    & + (1  -  \lambda) \log P_{\text{LM}}(\mathbf{Y} | \mathbf{h}_{\text{enc}}) \big\},
\end{aligned}
\label{eq:joint_decode}
\end{equation}
where $\lambda \in [0, 1]$ is a AM decoding weight. Here, $P_{\text{AM}}$ represents the acoustic probability predicted by the CROTTC branch, and $P_{\text{LM}}$ is the contextual probability evaluated autoregressively by $\text{Dec}(\cdot)$. The IF-MDD model is optimized via a multi-task objective:
\begin{equation}
    \mathcal{L}_{\text{total}} = \omega_{1} \mathcal{L}_{\text{AM}} + (1 - \omega_{1}) \mathcal{L}_{\text{LM}} + \omega_{2} (\mathcal{L}_{\text{pos}} + \mathcal{L}_{\text{type}}) + \omega_{3} \mathcal{L}_{\text{ga}},
\end{equation}
where $\mathcal{L}_{\text{ga}}$ is the guided-attention loss \cite{tachibana2018efficiently} ensuring monotonic fusion in $\text{FuN}$. Hyperparameters $\omega_{1}, \omega_{2}, \omega_{3}$ are set to 0.3, 1.0, and 10.0, respectively.

\section{Leveraging LLMs to Investigate Canonical Information Effect on MDD}
\label{sec:LLMMDD}
To empirically investigate the impact of explicit canonical information and strong linguistic priors, this section introduces an LLM-based architecture, denoted as \textbf{LLM-MDD}. As illustrated in Fig.~\ref{fig:LM2LLM}, we replace the conventional lightweight LM with an open-source LLM, which serves as a high-capacity linguistic processor for the acoustic embeddings. 
We adopt an LLM for this canonical information analysis for two primary reasons. First, LLMs inherently possess massive language modeling capabilities; therefore, leveraging an LLM to replace a lightweight LM itself represents a significant amplification of linguistic priors. Second, multimodal LLMs natively support the direct integration of text (canonical prompts) and speech (acoustic embeddings). This allows us to isolate and strictly verify the exact impact of various canonical injection strategies, while entirely bypassing complex and task-specific fusion mechanisms.

\subsection{LLM-MDD Framework}
Inspired by the SLAM-LLM architecture~\cite{ma2026slam}, our LLM-MDD system formulates the MDD task as mapping acoustic features and canonical cues to a perceived phoneme sequence. The architecture comprises three primary components: a pretrained AM, a trainable projector, and a backbone LLM.

During training, the AM remains frozen. We optimize the LLM via Low-Rank Adaptation (LoRA)~\cite{hu2022lora} and fully fine-tune the projector to align acoustic features to the LLM's embedding space. The system is optimized using standard cross-entropy loss against ground-truth perceived phoneme sequences. By injecting these massive linguistic priors, we aim to evaluate whether the emergent capabilities of LLMs can genuinely outperform traditional lightweight decoders in diagnosing subtle mispronunciations.

\subsection{Canonical Information Injection via Prompting}
\label{sec:LLMPrompting}
To explore how explicit canonical information influences the model's diagnostic accuracy, we design four distinct prompting strategies for LLM-MDD:

\noindent \textbf{Basic:} A standard SLAM-style sequence where the speech features are followed by the perceived phonemes: \texttt{[Prompt]\seqsplit{<speech><BOS><perc\_phns><EOS>}}.

\noindent \textbf{Canonical Injection:} Following the implementation in \cite{wu2024prompting}, the canonical phoneme sequence is appended to a general prompt: \texttt{[Prompt]\seqsplit{<cano\_phns><speech><BOS><perc\_phns><EOS>}}.

\noindent \textbf{Interleaved Potential Pronunciation:} Following the implementation in \cite{wu2025integrating, Song2025PhonemeControlledLW}, we interleave canonical phonemes with potential pronunciation (PP) alternatives: \texttt{[Prompt]\seqsplit{<cano\_phns\_with\_PP><speech><BOS><perc\_phns><EOS>}}. In both the training and inference stages, these PP candidates are uniformly assigned to each canonical phoneme. This design ensures that the model receives no prior clues regarding the specific \textit{location} or \textit{type} of error.

\noindent \textbf{Potential Pronunciation with Error Position:} Unlike the uniform assignment, we only provide PP candidates for phonemes explicitly annotated as mispronounced: \texttt{[Prompt]\seqsplit{<cano\_phns\_with\_PP\_err\_pos><speech><BOS><perc\_phns><EOS>}}. Since the mispronunciation position is pinpointed, the task reduces to a forced-choice mispronunciation diagnosis. Although this non-causal prompting style is impractical for real-world applications, this configuration is designed to test the performance upper bound of LLMs—specifically, whether they prioritize the canonical or the acoustic features in MDD.

\section{Experimental Evaluation}
This section comprehensively evaluates our CROTTC-IF framework against conventional dictation- and text-prompting-style baselines, alongside an empirical analysis of canonical priors using LLM-MDD.
\subsection{Datasets}
\label{sec:exp_setup}
\label{sec:dataset}
Table~\ref{tab:datasets_l2} summarizes the L2 English datasets utilized in our experiments. The most widely recognized among these is L2-ARCTIC, which standardized test set with six speakers (NJS, TLV, TNI, TXHC, YKWK, and ZHAA). Besides, we incorporated two supplementary datasets: Speechocean762 (SO762) and ERJ. SO762 was originally designed for pronunciation assessment, assigning each phoneme a score ranging from 0.0 to 2.0. Phonemes scoring below 0.5 are identified as mispronunciations, and their corresponding expert-annotated phonetic realizations are provided as ground truth. ERJ is a Japanese-accented L2-English corpus that serves as an out-of-domain (OOD) dataset to evaluate generalization ability. For vocabulary preparation, we adopted a 44-unit ARPAbet-style inventory (39 phonemes plus 5 special tokens \texttt{<bos>}, \texttt{<eos>}, \texttt{<sil>}, \texttt{<error>}, and \texttt{<blank>}).
\subsection{Implementation Details}
\label{sec:training_config}
\textbf{Acoustic Model (AM).} \label{sec:config_am}
For the AM backbone described in Sec.~\ref{sec:CROTTC}, we utilize WavLM Large~\cite{chen2022wavlm} followed by a 2-layer Conformer~\cite{gulati2020conformer} with a kernel and stride size of 3. The hidden dimension is set to 384. For the Consistency Regularization, we apply time warping with a factor of 80 alongside time and frequency masking following ~\cite{yaocr}. Specifically, we use a maximum of 3 mask blocks with a masking ratio $\rho \in (0.1, 0.3]$. In the time domain, the minimum masking length is enforced to ensure the CR mechanism leverages neighboring context rather than acting as simple data augmentation.

\noindent \textbf{Language Model (LM).} As introduced in Sec.~\ref{sec:IFMDD}, our LM comprises a 2-layer Transformer decoder with a hidden size of 384. We employ a two-stage training strategy: (1) initial joint CTC/Transformer training, followed by (2) substituting the CTC AM with the pretrained CROTTC AM and fine-tuning the LM until convergence. This approach resolves convergence issues arising from the absence of blank tokens in CROTTC, which otherwise provide crucial implicit segmental boundaries. The AM/LM training weight $\omega_{1}$ is set to 0.5, and the effect of the decoding weight is analyzed in Sec.~\ref{sec:lm_effect}.
 
\noindent \textbf{Auxiliary Teacher Network.} The teacher network in Sec.~\ref{sec:teacher} comprises two parallel branches: an encoder-side $\text{FuN}_{\text{enc}}(\cdot)$ and a decoder-side $\text{FuN}_{\text{dec}}(\cdot)$, both implemented with 2-layer Transformer decoders with a hidden size of 384. Notably, the 1D CNN down-sampler (factor=4) is integrated into $\text{FuN}_{\text{enc}}(\cdot)$ for alignment improvement. The shared CNN trunk from the mispronunciation head has a dimension of 128, followed by a 64-dim CNN branch for error-position detection and a linear branch for error type detection, respectively.
 
\noindent \textbf{LLM-MDD.} For the modules mentioned in Sec.~\ref{sec:LLMMDD}, we adapted AMs fine-tuned with both CTC and CROTTC. The projector is a 1D-CNN with a down-sampling factor of 4. We employed LLaMA-3.2-1B-Instruct~\cite{grattafiori2024LLaMA} and Qwen3-4B-Thinking-2507~\cite{qwen3technicalreport} as backbone LLMs, with rank=16 for LoRA fine-tuning. For the potential pronunciation prompting, we conducted a corpus-level statistical analysis, where the top 5 alternatives were selected as the PP candidates.
 
\noindent \textbf{Miscellaneous.} The AMs were trained for 300 epochs with a batch size of 32; learning rates were set to $10^{-5}$ for the SSL backbone and $3 \times 10^{-4}$ for the following modules. The LM and Teacher Network were trained for 200 epochs (batch size 32, learning rate $2 \times 10^{-4}$). LLM training was limited to 20 epochs (batch size 4, learning rate $2 \times 10^{-4}$). During inference, both the LM and LLMs utilized beam search decoding with a beam size of 10 and a temperature of 1.1. All experiments were conducted on a single NVIDIA GH200 GPU. Model selection was determined by the optimal F1-score on the validation set.
 
\begin{table}[t] 
\centering
\caption{Overview of the L2-ARCTIC, ERJ, and speechocean762 datasets.}
\vspace{-2mm}
\label{tab:datasets_l2}
\scriptsize
\renewcommand{\arraystretch}{1.1}
\setlength{\tabcolsep}{2.0pt}
\begin{tabular}{@{}llccccc@{}}
\toprule
\textbf{Dataset} & \textbf{L1} & \textbf{Volume} & \textbf{Train Utt.} & \textbf{Val Utt.} & \textbf{Test Utt.} & \textbf{Test Spk.} \\
\midrule
L2-ARCTIC \cite{zhao2018l2arctic}& Mixed\footnotemark & 3.66h & 2429 & 268 & 900 & 6 \\
SO762\cite{zhang2021speechocean762}& Mandarin & 5.58h & 2250 & 250 & 2500 & 125 \\
 ERJ\cite{ERJ} & Japanese\footnotemark & 0.99h & 674 & 74 & 144 &8 \\
\bottomrule
\end{tabular}
\vspace{-6mm}
\end{table}
\addtocounter{footnote}{-1} 
\footnotetext{Hindi, Korean, Mandarin, Spanish, Arabic, and Vietnamese.}
\stepcounter{footnote}
\footnotetext{Only expert-annotated speech were utilized as described in \cite{makino2012english}.}
\subsection{Evaluation metrics}

\begin{figure*}[!t]
	\centering
	\includegraphics[width=\textwidth]{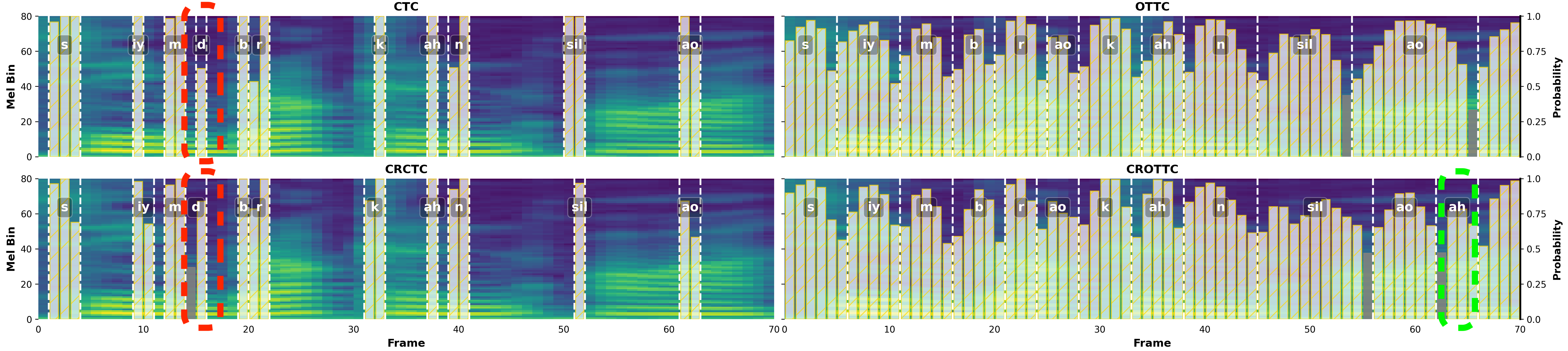}
\vspace{-4mm}
\caption{Comparison of frame-level probability distributions across different AMs (blank tokens omitted). The perceived phoneme sequence is \textbf{/s iy m (\textcolor{red}{d}) b r aa k ah n sil ao \textcolor[HTML]{34A853}{ah}/}, corresponding to the utterance \textbf{``seemed broken or'' $\rightarrow$ ``seem broken or-ah''}. While the CTC-based method hallucinated the \textcolor{red}{\textbf{/d/}} and omitted the \textcolor[HTML]{34A853}{\textbf{/ah/}}, our \text{CROTTC} faithfully captures the actual phonetic realization.}
	\label{fig:ctc_compare}
\vspace{-3mm}
\end{figure*}
We followed the hierarchical evaluation protocol of~\cite{7752846}. Detection was measured by true acceptance (TA), true rejection (TR), false acceptance (FA), and false rejection (FR), while diagnosis was evaluated with correct diagnosis (CD) and error diagnosis (ED) as illustrated in Table~\ref{tab:mdd_metrics}. The main performance is then summarized with the F1 score, treating TR as the positive class:
\vspace{-0.50\baselineskip} 
\begin{align*}
\text{P} &= \frac{TR}{TR + FR}, &
\text{R} &= \frac{TR}{TR + FA}, &
\text{F1} &= \frac{2 \text{P}\text{R}}{\text{P} + \text{R}}.
\end{align*}
We also report phoneme error rate (PER) and Correction Rate (COR) as a measure of recognition accuracy:
\vspace{-0.50\baselineskip} 
\begin{align*}
\text{PER} &= \frac{S + D + I}{N},&
\text{COR} &= 1-\frac{S + D}{N} ,&
\end{align*}
where $S$, $D$, and $I$ denote the numbers of substitutions, deletions, and insertions, and $N$ is the total number of \textbf{perceived phonemes}.

\begin{table}[h!]
\vspace{-4mm}
\centering
\caption{An illustration of MDD metrics.}
\vspace{-2mm}
\label{tab:mdd_metrics}
\scriptsize
\renewcommand{\arraystretch}{1.0}
\begin{tabular}{|l|c|c|c|c|c|}
\hline
\textbf{Canonical phoneme} & s & p & iy & k & t \\ \hline
\textbf{Perceived phoneme}& {\color[HTML]{34A853} s}  & {\color[HTML]{EA4335} b}  & {\color[HTML]{34A853} iy} & {\color[HTML]{EA4335} g}  & {\color[HTML]{EA4335} d}  \\ \hline
\textbf{Predicted phoneme} & {\color[HTML]{34A853} s}  & {\color[HTML]{34A853} p}  & {\color[HTML]{EA4335} ih} & {\color[HTML]{EA4335} g}  & {\color[HTML]{EA4335} th} \\ \hline
\textbf{Evaluation result} & TA & FA & FR & \begin{tabular}[c]{@{}c@{}}TR\\ CD\end{tabular} & \begin{tabular}[c]{@{}c@{}}TR\\ ED\end{tabular} \\ \hline
\end{tabular}
\vspace{-3mm}
\end{table}
 
\subsection{Experimental evaluation on L2-ARCTIC}
\label{Sec:Exp_l2arctic}
Table~\ref{tab:mdd_l2_arctic} compares our proposed framework against recent representative MDD methods on the L2-ARCTIC benchmark. Notably, CROTTC-IF achieves a peak F1 score of 71.77\%, yielding the lowest FRR and highest COR among all baselines. In the following subsections, we systematically analyze how the AM, LM, and canonical injection strategies collectively drive this superior performance.

\begin{table}[!t]
\centering
\scriptsize
\caption{Performance comparison on L2-ARCTIC (\%). \textbf{Bold} and \underline{underlined} values indicate the best and second-best results among fair-comparison models, respectively. $\uparrow$/$\downarrow$: higher/lower is better.}

\label{tab:mdd_l2_arctic}
\vspace{-2mm}
\renewcommand{\arraystretch}{1.1}
\setlength{\tabcolsep}{1.5pt}
\begin{tabular}{@{}lcccccccc@{}}
\toprule
\multirow{2}{*}{\textbf{Model}} & \multicolumn{3}{c}{\textbf{Detection}} & \multicolumn{3}{c}{\textbf{Diagnosis}} & \multicolumn{2}{c}{\textbf{Recognition}} \\ 
\cmidrule(lr){2-4} \cmidrule(lr){5-7} \cmidrule(lr){8-9}
& \textbf{F1 $\uparrow$} & \textbf{P} $\uparrow$ & \textbf{R}$\uparrow$ & \textbf{FRR} $\downarrow$ & \textbf{FAR} $\downarrow$ & \textbf{EDR} $\downarrow$ & \textbf{PER $\downarrow$} & \textbf{COR} $\uparrow$ \\

\midrule
\multicolumn{9}{c}{\textit{Dictation-Style Baselines}} \\
\midrule
MPL-MDD\cite{yang2022improving}   & 55.42 & 60.39 & 51.20 & 5.60 & 48.80 & 22.71 & 14.36 & - \\
RNN-T\cite{yue2023RNNT}           & 59.10 & 63.40 & 55.30 & 5.30 & 44.70 & -     & 15.47 & - \\
MV-w2v2\cite{el2023multi}         & 60.31 & 59.23 & 61.43 & -    & -     & -     & 14.13 & - \\
w2v2-CTC\cite{peng2021study}      & 60.44 & 62.86 & 58.57 & 5.70 & 41.80 & 29.28 & 16.20 & - \\
Meta-Learn \cite{metalearning}    & 61.45 & \textbf{91.60} & 46.24 & 29.75 & \textbf{8.40} & - & 42.25 & - \\

\midrule
\multicolumn{9}{c}{\textit{Text-prompting-Style Baselines}} \\
\midrule
Qwen2\cite{xie2026mispronunciation}& 50.60 & 71.62 & 39.12 & - & - & - & - & - \\
Qwen2-sub\cite{xie2026mispronunciation} & 55.00 & 63.27 & 48.64 & - & - & - & - & - \\
AEL w/o Pos.\cite{zhu2024pronunciation} & 56.33 & 58.36 & 55.00 & 6.55 & 45.00 & 25.72 & 14.81 & - \\
MDDGCN\cite{yan2023effective}     & 56.49 & 51.90 & 61.97 & 9.18 & 38.03 & 25.27 & - & - \\
Peppanet\cite{peppanet}           & 56.81 & 64.53 & 51.38 & 9.61 & 36.47 & 25.88 & - & - \\
PG-MDD\cite{lin2024pg}            & 60.10 & 60.06 & 60.15 & 6.65 & 39.94 & 23.13 & \underline{13.92} & - \\
TG+Contrast.\cite{peng2022text}   & 61.75 & 62.12 & 61.38 & 6.19 & 38.62 & 28.92 & - & - \\
\midrule
\multicolumn{9}{c}{\textit{Frame-wise MDD Baselines}} \\
\midrule
Joint-align\cite{lin2022phoneme}  & 63.04 & \underline{77.12} & 53.31 & - & - & - & - & - \\
PER-MDD\cite{tu2025mispronunciation}& \underline{69.60} & 71.78 & \textbf{67.56} & 4.43 & \underline{32.44} & 37.77 & 104.08 & \underline{90.42} \\
\midrule
\multicolumn{9}{c}{\textit{Proposed Methods (AM)}} \\
\midrule
\textbf{OTTC}         & 63.18 & 66.36 & 60.29 & 5.14 & 39.71 & 22.12 & 18.07 & 89.96 \\
 \textbf{CROTTC}                            & 62.39& 69.70& 56.47& 4.13& 43.53& 22.06 & 17.48 &90.29 \\
  \midrule
 \multicolumn{9}{c}{\textit{Proposed Methods (LM \& LLM)}} \\
  \midrule
\textbf{CTC-IF}              & 58.37 & 61.81 & 55.29 & 5.75 & 44.71 & \textbf{19.98} & \textbf{13.72} & 88.34\\
\textbf{CROTTC-IF}           & \textbf{71.77} & 76.94 & \underline{67.24} & \textbf{3.39} & 32.76 & 27.47 & 46.52 & \textbf{92.42} \\
 \textbf{CROTTC-LLaMA}                      & 56.87 & 54.81 & 59.08 & 8.20 & 40.92 & 21.98 & 15.85 &86.55 \\
\textbf{CROTTC-Qwen}                       & 55.19 & 58.00 & 52.64 & 6.42 & 47.36 & 23.80 & 15.42 &86.78 \\

\bottomrule
\end{tabular}
\vspace{-2mm}
\raggedright
\vspace{-3mm}
\end{table}

\subsubsection{Escaping the Acoustic Trap: Ablation on CROTTC}
\begin{table*}[t]
\centering
\caption{Comprehensive ablation studies on L2-ARCTIC. (a) Acoustic model variations. (b) Indirect Fusion components. (c) Canonical prompting strategies on LLM-MDD.}
\label{tab:combined_ablations}
\vspace{-3mm}

\resizebox{\textwidth}{!}{%
\renewcommand{\arraystretch}{1.1}
\setlength{\tabcolsep}{3pt} 

\begin{tabular}[t]{@{}c@{}}
\textbf{(a) Ablation of CROTTC AM} \\[-0.2mm]
\label{tab:AM_ablation}
\begin{tabular}[t]{@{}lccccccc@{}}
\toprule
\textbf{Method} & \textbf{F1}$\uparrow$ &\textbf{P}$\uparrow$ & \textbf{R}$\uparrow$ & \textbf{FRR}$\downarrow$ & \textbf{FAR}$\downarrow$ & \textbf{PER}$\downarrow$ &\textbf{INS}$\downarrow$ \\
\midrule
CTC     & 57.89 & 59.40 & 56.45 & 6.49 & 43.55 & 14.45 & 2.21 \\
CRCTC   & 58.45 & 61.40 & 55.78 & 5.90 & 44.22 & \textbf{14.01} & \textbf{2.05} \\
OTTC    & \textbf{63.18} & 66.36 & \textbf{60.29} & 5.14 & \textbf{39.71} & 18.35 & 8.31 \\
CROTTC  & 62.39& \textbf{69.70} & 56.47 & \textbf{4.13} & 43.53 & 17.48 & 7.76 \\
\bottomrule
\end{tabular}
\end{tabular}%
\hspace{1em}
\begin{tabular}[t]{@{}c@{}}
\textbf{(b) Ablation of IF LM} \\[-0.2mm]
\label{tab:IF_ablation}
\begin{tabular}[t]{@{}lccccc@{}}
\toprule
\textbf{Method} & \textbf{F1}$\uparrow$ & \textbf{P}$\uparrow$ & \textbf{R}$\uparrow$ &\textbf{FAR}$\downarrow$ & \textbf{EDR}$\downarrow$ \\
\midrule
CTC-IF                      & \textbf{58.37} & \textbf{61.81} & 55.29 & 44.71 & \textbf{19.98} \\
\quad w/o $\text{FuN}_{\text{Enc}}$ & 57.33 & 60.29 & 54.65 & 45.35 & 21.74 \\
\quad w/o $\text{FuN}_{\text{Dec}}$ & 56.51 & 59.42 & 53.86 & 46.14 & 22.10 \\
\quad w/o Error Head        & 57.56 & 59.54 & \textbf{55.71} & \textbf{44.29} & 22.28 \\
CTC-LM    & 54.95 & 59.23 & 51.25 & 48.75 & 22.60 \\
\bottomrule
\end{tabular}
\end{tabular}%
\hspace{1em}
\begin{tabular}[t]{@{}c@{}}
\textbf{(c) Ablation of LLM Canonical Prompts}\\[-0.2mm]
\label{tab:LLM_ablation}
\begin{tabular}[t]{@{}lccccccc@{}}
\toprule
\textbf{Method} & \textbf{F1}$\uparrow$& \textbf{P}$\uparrow$& \textbf{R}$\uparrow$& \textbf{FRR}$\downarrow$& \textbf{FAR}$\downarrow$& \textbf{EDR}$\downarrow$& \textbf{PER}$\downarrow$ \\
\midrule
CTC-LLaMA      & \underline{55.16} & 54.76 & \underline{55.57} & 7.73 & \underline{44.43} & \underline{24.17} & 16.23 \\
CROTTC-LLaMA   & \textbf{56.87} & \underline{54.81} & \textbf{59.08} & 8.20 & \textbf{40.92} & \textbf{21.98} & 15.85 \\
\quad w/ cano.           & 40.52 & \textbf{68.22} & 28.83 & \textbf{2.26} & 71.17 & 32.56 & \textbf{13.55} \\
\quad w/ PP              & 42.63 & 54.11 & 35.18 & \underline{5.02} & 64.82 & 35.15 & \underline{14.91} \\
\quad \textcolor{gray}{w/ pos. (oracle)}& \textcolor{gray}{\textbf{91.78}} & \textcolor{gray}{\textbf{95.02}} & \textcolor{gray}{\textbf{88.74}} & \textcolor{gray}{\textbf{0.78}} & \textcolor{gray}{\textbf{11.16}} & \textcolor{gray}{24.72} & \textcolor{gray}{\textbf{5.04}} \\
\bottomrule
\end{tabular}
\end{tabular}%
}
\vspace{-3mm}
\end{table*}

Table~\ref{tab:combined_ablations}(a) details the performance differences between CTC- and OTTC-based AMs. Compared to standard CTC, CRCTC yields improvements in both F1 score and PER. For OT-based methods, the F1 score exhibits a significant improvement, increasing from 57.89\% to 63.18\%. Concurrently, the PER also rises from 14.45\% to 18.35\%, with the most notable change being a surge in insertion errors from 2.21\% to 8.31\%. The integration of CR effectively mitigates these insertion errors down to 7.76\% and reduces the FRR to 4.13\%. In MDD, FRR is a critical usability metric; incorrectly flagging a student's correct pronunciation as an error can lead to learner frustration and undermine the pedagogical value of the system. Therefore, we select CROTTC as our primary AM, albeit with a slight trade-off in the overall F1 score.

To understand these acoustic mechanisms, we illustrate the frame-wise probability distributions of argmax predictions across different loss functions in Fig.~\ref{fig:ctc_compare}. Unlike CTC, OT-based methods assign non-blank tokens to every frame, which increases inference sensitivity but also leads to higher insertion errors during greedy decoding. By incorporating CR, the model effectively captures more subtle pronunciation deviations, successfully inheriting the alignment flexibility of OTTC while maintaining the recognition stability.

\subsubsection{Escaping the Linguistic Trap: Implicit Knowledge Transfer via IF}
\begin{figure}[t]
	\centering
	\includegraphics[width=0.46\textwidth]{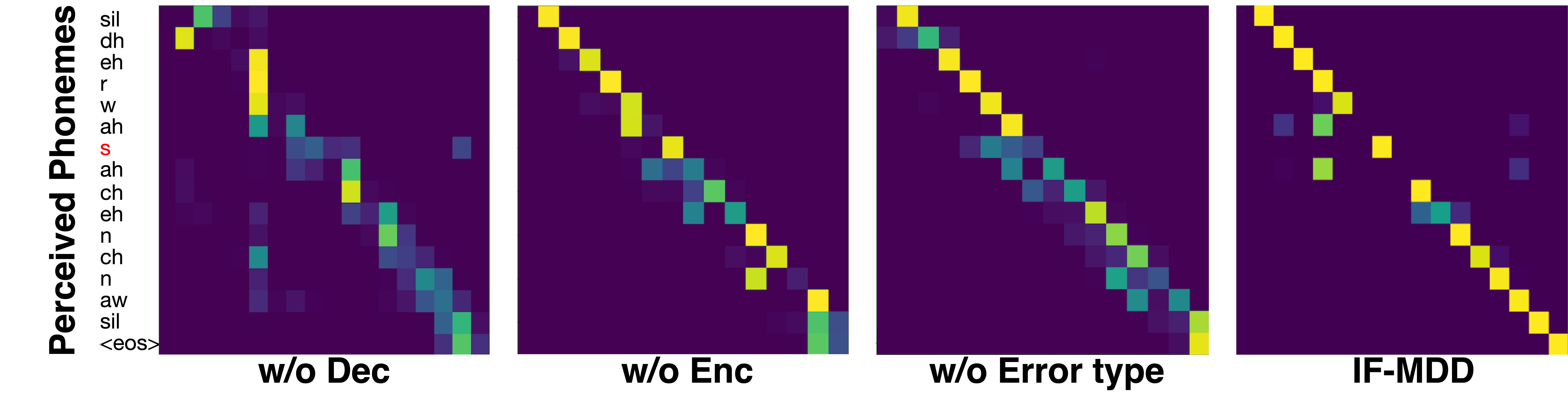}
\vspace{-3mm}
        \caption{Attention heatmaps of the Fusion Network under different ablation
conditions. The x-axis denotes the memory, and the y-axis denotes
the canonical-phoneme embeddings.}
	\label{fig:attn}
\vspace{-5mm}
\end{figure}

Table~\ref{tab:combined_ablations}(b) presents the ablation of the auxiliary teacher network, where CTC served as the AM backbone. Removing either $\text{FuN}_{\text{Enc}}$ or $\text{FuN}_{\text{Dec}}$ leads to a clear drop in the F1 score by 1.04\% to 1.86\%, demonstrating the effectiveness of multi-aspect monitoring in our Indirect Fusion strategy. Ablating the error-type classification head yields a smaller but consistent decrease in F1 alongside an increase in EDR, suggesting that this multi-view design helps to regularize the mispronunciation representations. Visualizations in Fig.~\ref{fig:attn} display the cross-attention heatmaps of the fusion network's last layer: removing these components produces blurrier, less-peaky patterns with reduced diagonal dominance, reflecting the degraded representation quality of the mispronunciation cues. Furthermore, without the teacher network's knowledge transfer, the model devolves into a standard CTC/Transformer ASR architecture (CTC-LM), resulting in a significant decrease in the F1 score. This proves that uncritically adapting general LM-based ASR models to the standalone MDD task generates suboptimal diagnostic results.

\subsubsection{The Danger of Explicit Priors: Canonical Injection with LLMs}
Table~\ref{tab:combined_ablations}(c) presents our investigation into the effect of strong canonical priors on MDD performance. First, compared to CTC-LLaMA, CROTTC-LLaMA yields better detection performance, boosting the F1 score from 55.16\% to 56.87\%. This supports our hypothesis that the sparsity inherent in CTC causes MDD models to overlook transient mispronunciation cues. Second, by varying the canonical information injected into the prompt, we observe that explicitly providing canonical targets is highly detrimental to MDD, resulting in a drastic drop in the F1 score to 40.52\%. While prompting with Potential Pronunciations (PP) offers the LLM slightly more flexibility, the performance recovery remains severely limited (F1 = 42.63\%), consistent with the observation in \cite{xie2026mispronunciation}. These observations strongly demonstrate the ``Linguistic Trap'' in MDD: strong textual priors bias the LLM toward over-correction, masking objective acoustic fidelity. Third, when prompting the LLM with both PP and explicit mispronunciation positions, the F1 score naturally sees a significant improvement to 91.78\%. However, a gap from perfect MDD remains: while FRR drops to 0.78\%, EDR and FAR remain high at 24.72\% and 11.16\%. This proves the fundamental bottleneck for multimodal LLMs in MDD is not merely detecting error positions, but also the inability to utilize the fine-grained acoustic resolution required for precise diagnosis. Even with explicit positional hints, the LLM struggles to accurately identify specific substitutions and stubbornly defaults to the textual prior. Therefore, we believe that if future research aims to utilize multimodal LLMs for MDD, developing more refined acoustic processing stages is essential.

\vspace{-0.55\baselineskip} 
\subsubsection{Balancing Acoustic Fidelity and Linguistic Context}
\label{sec:lm_effect}
\begin{figure}[t]
	\centering
	\includegraphics[width=0.45\textwidth]{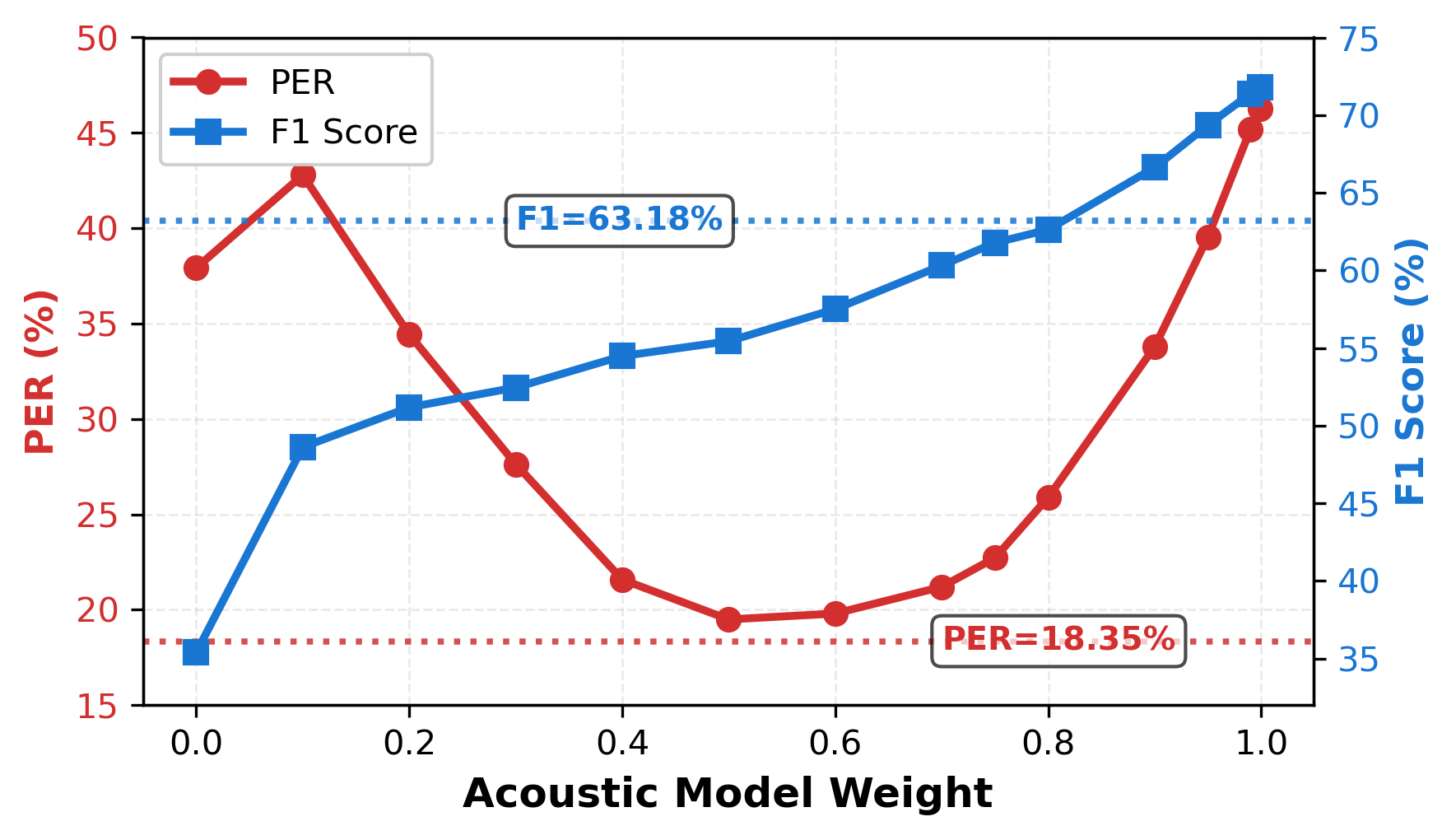}
\vspace{-3mm}

\caption{Trade-off between PER and F1 on the L2-arctic over varying $w_{\text{AM}}$ .}
	\label{fig:lm_effect}
\vspace{-4mm}
\end{figure}
As formulated in Eq.~\ref{eq:joint_decode}, CROTTC-IF performs decoding under a joint AM/LM strategy. In standard CTC/LM architectures for general speech recognition, a sweet spot for the AM weight (typically $\lambda \in [0.2, 0.5]$) is usually found to achieve the optimal Word Error Rate (WER). However, for the standalone MDD task, we hypothesize that a significantly higher AM weight is crucial for preserving detection sensitivity. To validate this, we conducted a spectrum-control study on the AM weight $\lambda$. 

Figure~\ref{fig:lm_effect} clearly demonstrates this divergence: while the PER reaches its optimum around $\lambda = 0.5$, the F1 score exhibits a completely different trajectory.When $\lambda = 0$, the F1 drops to 36.83\%, proving that relying solely on the LM's contextual priors leads to mispronunciation detection failure. As $\lambda$ increases, F1 shows a monotonically increasing trend, confirming our hypothesis that acoustic fidelity is fundamentally more critical than linguistic priors in MDD. Notably, when $\lambda > 0.8$, the joint decoding achieves superior performance compared to CROTTC AM alone, demonstrating that IF-MDD provides effective and complementary soft linguistic guidance\footnotemark{}.
\footnotetext{In MDD, PER is considered secondary to F1 and FRR\cite{cnn-rnn-ctc, tu2025mispronunciation}. High PER often reflects redundant insertions that do not degrade diagnostic accuracy, as evidenced by our robust COR at 92.42\%.}

\vspace{-0.55\baselineskip} 
\subsection{Generalization Study on SO762 and ERJ}
Table~\ref{tab:erj_results} presents our experimental results on two additional L2 datasets: SO762 and ERJ. For the zero-shot experiments, we directly evaluated the models trained on L2-ARCTIC (as described in Section~\ref{Sec:Exp_l2arctic}), and further adapted to the mentioned datasets for the fine-tuned settings. Compared to conventional baselines on SO762, CROTTC-IF achieves a highly competitive F1 score of 57.16\% with significantly better FAR and EDR, without requiring the explicit fine-grained phonetic score labels as used in  \cite{JAM, MuFFIN}. Furthermore, CROTTC-IF delivers outstanding fine-tuned and remarkably strong zero-shot results on ERJ, clearly demonstrating the robust out-of-domain (OOD) generalization of our prompt-free architecture.

\begin{table}[t]
\centering
\caption{Fine-tuned and zero-shot results on SO762 and ERJ.}
\label{tab:erj_results}
\vspace{-2mm}
\resizebox{\linewidth}{!}{%
\renewcommand{\arraystretch}{1.1}
\setlength{\tabcolsep}{3pt} 
\begin{tabular}{@{}lllcccccc@{}}
\toprule
\textbf{Dataset} & \textbf{Cond.} & \textbf{Model} & \textbf{F1} $\uparrow$ & \textbf{P} $\uparrow$ & \textbf{R} $\uparrow$ & \textbf{FRR} $\downarrow$ & \textbf{FAR} $\downarrow$ & \textbf{EDR} $\downarrow$ \\ 
\midrule
\multirow{6}{*}{SO762} & \multirow{6}{*}{Fine-tuned}
  & Ryu2023 \cite{ryu23_interspeech} & 41.50 & 26.90 & \textbf{91.60} & - & - & - \\
& & JAM \cite{JAM}      & 45.01 & \underline{61.10} & 34.76 & \textbf{0.58} & 64.32 & \underline{45.23} \\
& & MuFFIN \cite{MuFFIN}& \textbf{67.98} & \textbf{67.60} & 68.37 & \underline{1.01} & 31.63 & 58.82 \\ 
\cmidrule(l){3-9}
& & CTC-IF       & 46.68 & 31.86 & 87.23 & 6.32 & 12.77 & 50.30 \\
& & CROTTC       & 51.86 & 36.81 & 87.67 & 5.09 & \underline{12.32} & 57.40 \\
& & CROTTC-IF    & \underline{57.16} & 41.90 & \underline{89.92} & 8.63 & \textbf{6.53}  & \textbf{10.08} \\ 
\midrule
\multirow{6}{*}{ERJ} & \multirow{3}{*}{Fine-tuned} 
  & CTC-IF       & 83.98 & 83.91 & 84.06 & 9.70& 15.94 & \underline{26.60} \\
& & CROTTC       & \underline{85.79} & \underline{85.79} & \underline{85.79} & \underline{8.63} & \underline{14.21} & 29.09 \\
& & CROTTC-IF    & \textbf{89.27} & \textbf{89.12} & \textbf{89.43} & \textbf{6.63} & \textbf{10.57} & \textbf{25.78} \\ 
\cmidrule(lr){2-9}
& \multirow{3}{*}{Zero-shot} 
  & CTC-IF       & 67.13 & 79.57 & 58.06 & - & 41.94 & 54.03 \\
& & CROTTC       & \underline{69.17} & \underline{83.37} & \underline{59.10} & \underline{7.16} & \underline{40.90} & \underline{46.92} \\
& & CROTTC-IF    & \textbf{78.44} & \textbf{87.26} & \textbf{71.23} & \textbf{6.32} & \textbf{28.77} & \textbf{44.77} \\ 
\bottomrule
\end{tabular}%
}
\vspace{-4mm}
\end{table}
\section{Iqra'Eval2 Challenge}
\begin{table}[!t]
\centering
\caption{Overview of the Iqra'Eval2 datasets.}
\vspace{-2mm}
\label{tab:datasets}
\scriptsize
\renewcommand{\arraystretch}{1.1}
\setlength{\tabcolsep}{1.0pt}
\begin{tabular}{@{}lccc@{}}
\toprule
\textbf{Dataset} & \textbf{Volume} & \textbf{Error Type} & \textbf{Description} \\
\midrule
\texttt{Iqra}   & $\sim$79h & None& Native speakers; golden reference \\
\texttt{TTS}    & $\sim$80h & Synthetic      & TTS-injected mispronunciations \\
\texttt{Extra}  & $\sim$2h  & Real errors    & Real-world human mispronunciations \\
\bottomrule
\end{tabular}
\vspace{-3mm}
\end{table}
The Iqra'Eval2 Challenge \cite{kheir2025unifiedbenchmarkarabicpronunciation} provides an ideal arena for our framework. Unlike standard L2 English benchmarks, Qur'anic recitation is governed by \textit{Tajweed} rules\footnote{A strict set of phonetic rules governing Qur'anic recitation, distinct from Modern Standard Arabic (MSA).}, demanding precise modeling of nuanced articulatory deviations. Based on the empirical findings from our L2 English experiments, we deployed our CROTTC-IF architecture for this challenge. 

Table~\ref{tab:datasets} summarizes the challenge's datasets. Among the provided corpora, we exclusively utilized the \texttt{TTS} and \texttt{extra} datasets. Specifically, we applied the \texttt{TTS} corpus for the AM's baseline training, followed by fine-tuning on the \texttt{extra} corpus, which contains genuine human pronunciation errors. For the \texttt{extra} corpus, we randomly allocated 10\% of the utterances as a local test set, with the remainder used for training and validation. For vocabulary preparation, we followed \cite{halabi-wald-2016-phonetic}, utilizing an inventory of 67 Arabic phonemes\footnote{We omitted ``$<<$", the geminated glottal stop, as it is not represented in our selected training data.} plus 4 special tokens (\texttt{<bos>}, \texttt{<eos>}, \texttt{<sil>}, and \texttt{<blank>}), resulting in a vocabulary size of 71. Other training configurations remain consistent with those detailed in Sec~\ref{sec:training_config}.
During inference, we applied a high acoustic weight ($\lambda = 0.9$) during joint decoding as discussed in Sec~\ref{sec:lm_effect}. As shown in Table~\ref{tab:leaderboard_results}, this principled approach yielded outstanding results. Operating entirely without explicit canonical prompts, our CROTTC-IF system ranked 2nd on the official Iqra'Eval2 leaderboard, achieving a highly competitive F1-score of 71.70\% and a remarkably low PER of 3.72\%. These results further demonstrate our core theory: decoupling acoustic modeling from explicit canonical priors can provide a robust and objective paradigm for real-world MDD tasks.

\begin{table}[t]
  \centering
  \scriptsize
  \setlength{\tabcolsep}{1.5pt}
  \caption{Performance of CROTTC-IF on the Iqra'Eval2 leaderboard. }
  \vspace{-3mm}
  \label{tab:leaderboard_results}
  \begin{tabular}{@{}lcccc@{}}
    \toprule
    \textbf{Model} &  \textbf{F1$\uparrow$}&\textbf{Pre. $\uparrow$}& \textbf{Rec.$\uparrow$}& \textbf{PER $\downarrow$} \\
    \midrule
 3rd-team& 71.57& 67.69& 75.93&4.05\\
 \midrule
    CROTTC &  68.77&70.07 & 67.52 & 4.11 \\
   \quad w/ IF ($\lambda = 0.3$) &  70.72&72.67 & 68.89 & 3.82 \\
 \textbf{\quad w/ IF ($\lambda = 0.9$) }& \textbf{71.70}& \textbf{73.25} & \textbf{70.20} &\textbf{3.72} \\
  \midrule
  1st-team& 72.01& 74.16& 69.98&3.65\\
 \bottomrule
  \end{tabular}
\vspace{-6mm}
\end{table}

\section{Conclusion and Future Work}
In this paper, we proposed \textbf{CROTTC-IF}, a canonical prompt-free framework designed to overcome the inherent acoustic and linguistic bottlenecks in MDD. By integrating a dedicated acoustic frontend with an implicit knowledge transfer language model, our architecture achieves robust, state-of-the-art performance across diverse scenarios. Furthermore, through comprehensive LLM prompting experiments, we empirically demonstrated that uncritically injecting explicit canonical priors severely compromises acoustic fidelity. This highlights the critical need for careful and principled designs when incorporating canonical information. By successfully decoupling acoustic modeling from explicit textual priors, we believe this prompt-free paradigm establishes a highly objective and robust foundation for the MDD field. Future work will focus on further optimizing the detection-recognition trade-off and extending the framework to other CAPT applications.

\section{ Generative AI Use Disclosure}
Generative AI technology was employed strictly for minor grammatical corrections and stylistic improvements during the drafting process. The authors have verified all content and retain full responsibility for the accuracy and originality of this work.



\bibliographystyle{IEEEtran}
\bibliography{mybib}

\end{document}